\documentclass[10pt,twocolumn,superscriptaddresst]{revtex4-2}

\usepackage[english]{babel}
\usepackage{graphicx}
\usepackage{color}
\usepackage{caption}
\usepackage{upgreek}

\usepackage{textcomp}

\usepackage{enumitem,amssymb}
\newlist{todolist}{itemize}{2}
\setlist[todolist]{label=$\square$}
\usepackage{pifont}

\begin{document}

\title{Deterministic generation of large-scale photonic GHZ states utilizing spin echo in a solid-state quantum light source}

\author{
	Zu-En Su \textsuperscript{1,2,!}, Dan Cogan\textsuperscript{1,!}, Oded Kenneth\textsuperscript{1}, Shunfa Liu\textsuperscript{3}, Jin Liu\textsuperscript{2,3} and David Gershoni\textsuperscript{1,2,4*} \\%[1ex]
}

\affiliation{
	\textsuperscript{1} The Physics Department and the Solid State Institute, Technion-Israel Institute of Technology, Haifa 3200003, Israel \\
	\textsuperscript{2} Quantum Science Center of Guangdong-Hong Kong-Macao Greater Bay Area, Shenzhen 518045, Guangdong, China\\
	\textsuperscript{3} State Key Laboratory of Optoelectronic Materials and Technologies, School of Physics, Sun Yat-sen University, Guangzhou 510275, Guangdong, China\\
	\textsuperscript{4}Guangdong Technion-Israel Institute of Technology, Shantou 515063, Guangdong, China\\
	\textsuperscript{!}Equal contribution, \textsuperscript{*}Corresponding author: \href{mailto:dg@physics.technion.ac.il}{dg@physics.technion.ac.il}
}

\begin{abstract}
	Quantum technologies witness rapid contemporary developments transitioning from abstract scientific ideas and fundamental demonstrations, into systems of sufficient quality and scale, enabling real applications. Photonic quantum technologies are playing a pivotal role in these developments, since photons are robust, easily controlled, detect and measured, thus qualify as flying carriers of quantum information and entanglement distributers. The efforts to develop photonic quantum technologies benefit from the maturity of the classical optical technologies. It was shown that even a source of heralded three entangled photons is sufficient for building large scale quantum computers using otherwise classical optical elements. Here, we demonstrate a novel semiconductor quantum dot based device, which deterministically generate at GHz rates, indistinguishable single photons in a robust Greenberger-Horne-Zeilinger (GHZ) state of more than a dozen photons. The device utilizes novel natural spin echo concept, provided by the opposite precession directions of the negative and positive carriers' spins in the excited and ground levels of the periodically excited dot, respectively. The excitation rate is finely tuned to nullify the phase acquired by both spins' precession. Our record long GHZ state producing device can yet be feasibly improved to provide higher-rate, brighter and more robust deterministic GHZ source, significantly reducing the overhead requirements for photonic quantum technologies.
\end{abstract}

\maketitle
\global\long\def\ket#1{\left|#1\right\rangle }%
\global\long\def\im{\operatorname{Im}}%
\global\long\def\bra#1{\left\langle #1\right|}%

\section{Introduction} \label{sec1}

Photonic quantum technologies are essential for the development of quantum information processing, since photons are ideal flying qubits to be used for carrying quantum information and for entanglement distribution.
The maturity of the classical photonic industry \cite{psiquantum2025} further contributes to the rapid developments of these technologies.

Multiphoton entangled graph and cluster states form an essential resource for both future quantum computing and communication \cite{psiquantum2025,o2009photonic,wang2020integrated,pelucchi2022potential,wang2025scalable}. 
In quantum computing, the built-in entanglement in the multiphoton state enables measurement-based computation \cite{raussendorf2003measurement,briegel2009measurement}, 
and in communication the inherent redundancy of the multiphoton state can compensate for photon loss and for the probabilistic nature of Bell state measurements,
% (BSMs) \cite{xxx}, 
while the locality of the optical measurements can be used for memory-free quantum networking \cite{azuma2015all}. 

Entangled multiphoton states can be generated in various ways \cite{browne2005resource,kok2007linear,ghanbari2024optimization}, 
and conceptually it has been shown that a quantum light source of only three heralded entangled indistinguishable photons is sufficient for building a large multiphoton entangled state, using otherwise only linear optical elements \cite{Rudolph2017}.
Efforts to produce heralded entangled three photons in a GHZ state were recently reported \cite{chen2024heralded3,cao2024heralded3}, 
but the use of these heralded states for scaling up is a formidable task which requires a lot of overhead resources \cite{lu2007experimental,yao2012observation,wang2016experimental,zhong201812}.
A more promising approach is to generate much larger multiphoton states in a deterministic, rather than heralded,  manner,  by exploiting for example, spin-photon interaction in single quantum emitters, as first proposed and demonstrated by Lindner and Rudolph \cite{lindner2009} and Schwartz and coauthors \cite{Schwartz2016}, respectively. 
In this scheme, periodic and timed optical excitations of a confined spin result in the emission of a string of polarization-entangled photons, due to specific optical selection rules that link the photon polarizations to the spin states. 
Remarkable advances along this direction with a variety of single quantum emitters such as Rydberg superatoms \cite{yang2022sequential}, 
Rubidium atoms \cite{thomas2022}, 
and semiconductor quantum dots (QDs) \cite{Cogan2023,coste2023,Su2024,meng2024deterministic,laccotripes2025entangled} have been recently demonstrated. 
QDs, in particular, feature the highest photon generation rates, reduced footprints and compatibility with the contemporary microelectronic and photonic industries which enable potential chip integration \cite{lodahl2015interfacing,heindel2023quantum,zhou2023epitaxial,psiquantum2025}. 
Cogan, Su and coauthors \cite{Cogan2023,Su2024} successfully demonstrated continuous deterministic generations of 1D cluster states with an entanglement length of up to $\sim$10 qubits, and high indistinguishability 
as required for photon fusion operations and scaling up of the entangled multiphoton state \cite{paesani2023fusion}.

Here we demonstrate deterministic generation of photonic GHZ state of more than a dozen indistinguishable photons at a photon generation rate which exceeds 1 GHz. 
For achieving these record features we utilize the natural spin echo that the QD confined charge carriers provide. 
The electron spin in the excited state before the photon emission and the hole in the ground state after the emission, precess in opposite directions in the presence of magnetic field. 
We thus tune the excitation rate to exactly nullify the acquired average phase during both spin precessions. Remarkably, this optimal rate does not depend on the magnitude of the magnetic field, adding robustness and stability to the device operation. 

\section{Results}\label{sec2}

\begin{figure}
	\begin{centering}
		\includegraphics[width=1\columnwidth]{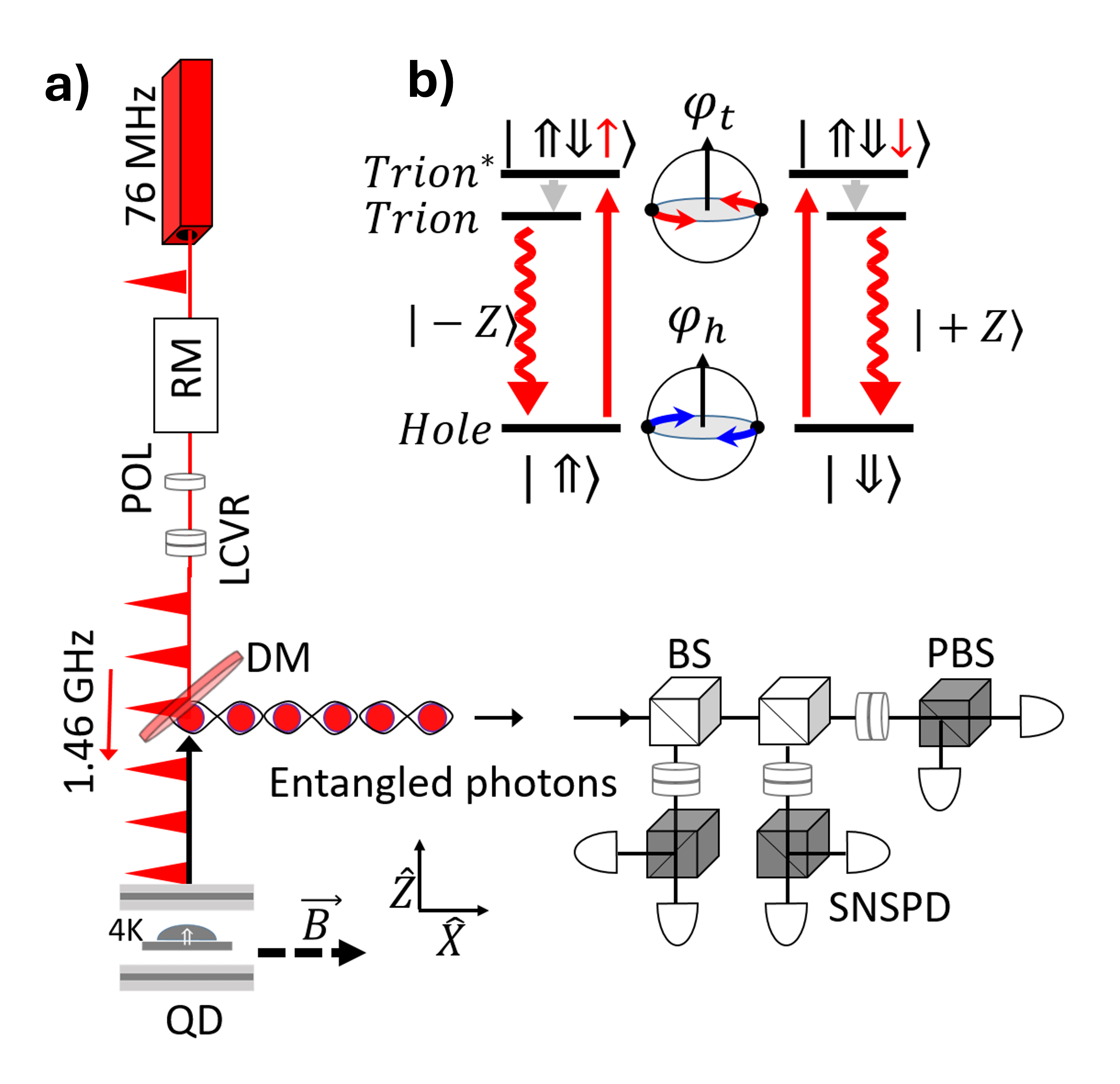}
		\par\end{centering}
	\caption{\label{fig: Setup} {\bf The Setup.} 
		({\bf a}) Schematic description of the QD-based device. 
		%A train of polarized laser $\pi-$pulses resonantly excite the QD resulting in deterministic emission of a single photon after each pulse. The photon polarization is entangled with the spin of the confined hole which precesses in the externally applied magnetic field (B) in Voigt configuration. 
		RM: rate multiplier, BS: beam splitter, PBS: polarizing beam splitter, SNSPD: superconducting nanowire single photon detector, LCVR: liquid crystal variable retarder, DM: dichroic mirror, and POL: polarizer.
		({\bf b}) Schematic of the optical $\Pi$-system between the heavy hole (HH) and the positive trion. %polarization selection rules enabling spin-photon entanglement. 
		$\ket{\pm Z}$ represents a right- (left-) hand circularly polarized photon.
		%The HH and the trion spins precess in opposite directions. 
		%due to the opposite signs of the electron and hole g-factors. 
		$\varphi_\mathrm{h(t)}$ marks the acquired phase due to the precession of the HH (trion in opposite direction) after (before) the radiative recombination.
	}
\end{figure}

%The device was essentially described previously \cite{Cogan2023,Su2024}. In brief, 
The heart of the device is an InAs/GaAs self-assembled semiconductor QD embedded in the center of a one-wavelength planar microcavity, formed by two AlAs/GaAs Bragg reflecting mirrors \cite{Cogan2023,Su2024} and maintained at ambient temperature of 4~K (Fig.~\ref{fig: Setup}a). 
%Figure 1

%The device is maintained at ambient temperature of 4~K. 
The QD confines a heavy-hole (HH) which is periodically excited optically using resonant 12~ps laser $\pi$-pulses \cite{Cogan2023}. Each laser pulse photogenerates an additional electron and HH pair, which together with the resident HH form a positive trion. The resonant excitation is tuned to an excited confined electronic level from which the electron relaxes to the ground trion level by emitting spin-preserving longitudinal-optical (LO) phonon \cite{Benny2011,Benny2012}. Since the LO phonon resonates with the electronic levels' energy separation \cite{Benny2014}, the relaxation is very fast. Its duration (about 5 ps) can be estimated from the spectral width of the absorption resonance of this optical transition (about 300 $\upmu$eV) and from analysis of the excitation intensity dependence of the trion PL \cite{Cogan2023}. 
The trion then radiatively recombines by emitting a single photon. 
%During the recombination the unpaired electron spin precesses in the magnetic field.
This quasiresonant ``p-shell'' excitation has the following advantages: i) It provides $\pi$-pulse like deterministic excitation. ii) The reflected laser light can be spectrally filtered without blinding the detectors. iii) The fast electron relaxation is comparable to the laser pulse temporal width, therefore it barely affects the excited state's jitter thus maintaining photon indistinguishability of about 0.95 \cite{Cogan2023,Su2024}.

Spin-photon entanglement is created 
%in this system 
by utilizing the polarization selection rules of the optical transition between the ground HH and the excited trion \cite{Cogan2018}, which form an optical $\Pi$-system (Fig.~\ref{fig: Setup}b). For example, if a $\pi$-area rectilinearly horizontally polarized ($\ket{H}=1/\sqrt{2}(\ket{Z}+\ket{-Z})$, where $\pm Z$ is right (left) circular polarization) laser pulse is applied to a hole in a coherent superposition, $\ket{\Psi^\mathrm{h}} = \alpha\ket{\Uparrow}+\beta\ket{\Downarrow}$, a trion state, $\ket{\Psi^\mathrm{t}} = \alpha\ket{\Uparrow\Downarrow\uparrow}+\beta\ket{\Downarrow\Uparrow\downarrow}$ is formed. When the trion recombines, an entangled spin-photon state $\ket{\Psi^\mathrm{h,p}}= \alpha\ket{\Uparrow,-Z}+\beta\ket{\Downarrow,+Z}$ is generated. 
A three-qubit spin-photon-photon GHZ state $\ket{\Psi^\mathrm{h,p_1,p_2}}= \alpha\ket{\Uparrow,-Z_1,-Z_2}+\beta\ket{\Downarrow,+Z_1,+Z_2}$ is generated by repeating the process again \cite{Cogan2023}. Spin-$N$-photon state is similarly generated by repeating the process $N$ times: 
\begin{equation} 
	\ket{\Psi^\mathrm{h,p_1,...,p_N}}=\alpha\ket{\Uparrow}\ket{-Z}^{\otimes N}+\beta\ket{\Downarrow}\ket{+Z}^{\otimes N} 
	\label{GHZalpha}
\end{equation}

In order to deterministically prepare a coherent superposition of the HH spin states, we apply magnetic field parallel to the microcavity plane. 
The Zeeman interaction lifts the degeneracy between HH spin direction parallel and antiparallel to the direction of the field ($x$). 
The energy separation between the HH eigenstates results in a phase evolution that can be described as a coherent spin precession 
in the $yz$ plane around the magnetic field direction. By timing the precession, following detection of a circularly polarized photon $\ket{\pm Z}$, which heralds the HH spin state $\ket{\Downarrow (\Uparrow)}$, the phase of the superposition can be accurately determined. 

Timed spin precession is important for generating photonic cluster states \cite{Schwartz2016,Cogan2023, Su2024}; however, for a GHZ state, it is actually counterproductive, since in the presence of precession the generated spin-multiphoton state is not represented by Eq. \ref{GHZalpha}.  
Luckily, it turns out that the confined HH and electron spins have opposite in-plane $g$-factor signs \cite{Cogan2022,Su2025}. 
As a result the excited trion and the ground state HH precess in opposite directions. This opposite spin precession allows one spin to compensate for the phase acquired by the other spin, simply by timing the excitation rate.

\begin{figure}
	\begin{centering}
		\includegraphics[width=1\columnwidth]{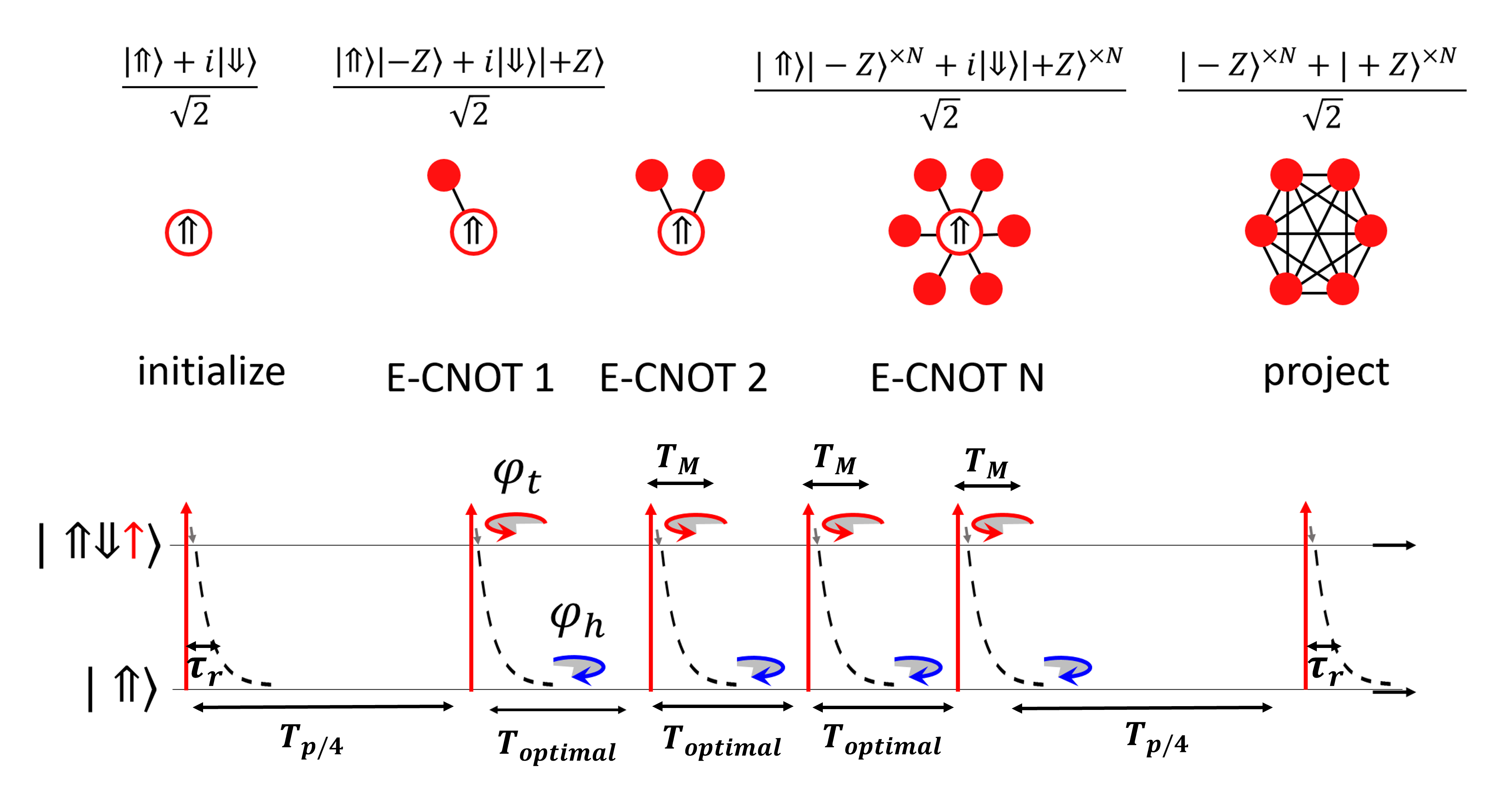}
		\par\end{centering}
	\caption{\label{fig: Method} {\bf The GHZ generation method.} 
		Schematic description of the temporal sequence of operations for generating the GHZ state. %The sequence is divided into three parts: 
		i) Hole spin initialization in a $\ket{\pm Y}$ state
		by projecting the first photon polarization on a circular base and timed hole spin precession of $T_\mathrm{p/4}$. 
		ii) Spin-$N$-photon entanglement by applying the E-CNOT gate $N$ times, and iii) Final HH spin projection onto a $\ket{\mp Y}$ state by detecting circularly polarized photon following $T_\mathrm{p/4}$ delayed excitation. 
		The spin-multiphoton state is diagrammatically depicted above each step. 
		The excitation period $T_\mathrm{optimal}$ is tuned to null the average acquired phase $\varphi_\mathrm{t}+\varphi_\mathrm{h}$. $\tau_\mathrm{r}$ is the radiative lifetime and $T_\mathrm{M}$ is the temporal width of the measurement photon detection window}
\end{figure}

Fig.~\ref{fig: Method} illustrates the method for generating and characterizing the photonic GHZ state. 
The HH spin is initialized by detecting a circularly polarized photon $\ket{\pm Z}$ and allowing the heralded HH spin $\ket{\Uparrow}$ ($\ket{\Downarrow}$) to precess for a quarter of a precession period ($T_\mathrm{p/4}$) to reach the desired coherent superposition $\ket{\Psi^\mathrm{h}_{\pm y}} = \sqrt{1/2}\left(\ket{\Uparrow}\mp i\ket{\Downarrow}\right)$. 
$N$ entangled photons in a GHZ state with the spin are then produced by repeated optical excitation using $H$-polarized $\pi$-pulses. 
The temporal separation between the pulses ($T_\mathrm{optimal}$) is chosen such that the fast electronic spin precession during the trion recombination time ($\tau_\mathrm{r}$), is compensated by timing the slow precession of the HH after the recombination and before the next excitation.
This way, by applying a train of $N$ laser pulses at a predefined rate a spin-$N$-photon GHZ state is generated: 
\begin{equation} 
	\ket{\Psi^\mathrm{h,p_1,...,p_N}}=1/\sqrt{2}(\ket{\Uparrow}\ket{-Z}^{\otimes N}\pm i\ket{\Downarrow}\ket{+Z}^{\otimes N}) 
	\label{GHZ}
\end{equation}
The protocol ends with projecting the HH spin on the $y$ direction ($\ket{\Psi^\mathrm{h}_{\mp y}}$)
%$\ket{\Psi^h_{\pm y}} = \sqrt{1/2}\left(\ket{\Uparrow_x}\mp i\ket{\Downarrow_x}\right)=\sqrt{1/2}\left(\ket{\Uparrow_z}\pm \ket{\Downarrow_z}\right)$$\ket{\Psi^h_{\pm y}} = \sqrt{1/2}\left(\ket{\Uparrow}\mp i\ket{\Downarrow}\right)$ 
by detecting a circularly polarized photon resulting from a pulsed excitation delayed by $T_\mathrm{p/4}$.
%a quarter precession period. 
This projection leaves behind $N$ entangled photons in a pure photonic GHZ state:
\begin{equation} 
	\ket{\Psi^\mathrm{p_1,...,p_N}}=1/\sqrt{2}(\ket{-Z}^{\otimes N}\pm\ket{+Z}^{\otimes N}) 
	\label{GHZphot}
\end{equation}

%Figure 2

The entangling gate applied $N$ times in Fig.\ref{fig: Method}  (E-CNOT gate) is composed of four steps:
i) An $H$-polarized laser $\pi$-pulse transforms the HH state into a trion state. ii) The trion precesses, adding a phase $\varphi_\mathrm{t}$ to the coherent superposition of its eigenstates. iii) The trion recombines during the radiative time $\tau_\mathrm{r}$, emitting a photon whose polarization is entangled with the confined HH spin. iv) The HH precesses, adding a phase $\varphi_\mathrm{h}$ before the next pulse arrives after an interval $T=\Gamma_\mathrm{exc}^{-1}$, where $\Gamma_\mathrm{exc}$ is the excitation repetition rate. 

The accumulated phase between two sequential photons in the GHZ state can be quantified by measuring their mutual degree of circular polarizations defined as $P_{Z}^{n}P_{Z}^{n+1}$ where $P_j^{i}$ represents photon $i$ projected on polarization direction $j$ \cite{toth2005detecting}.   
\begin{equation}\label{phase_NN+1}
	P_{Z}^{n}P_{Z}^{n+1}=A_\mathrm{dec}^n \cos (\varphi^n)
\end{equation} 
where $A_\mathrm{dec}^n\le1$ represents the decoherence of the spins during the two emissions and $\varphi^n=\varphi_\mathrm{h}^n+\varphi_\mathrm{t}^{n+1}$ is the accumulated phase between the two sequentially emitted photons.

Ideally, for the generation protocol to work, the acquired phase between every two sequential photons should vanish, (or be an integer product of $2\pi$ \cite{Peniakov2020}).
In the Method section we detailed the considerations for determining the optimal excitation rate ($\Gamma_\mathrm{exc}^\mathrm{opt}=T_\mathrm{optimal}^{-1})$, which nulls the average accumulated phase.
In addition we develop there a model for analyzing the dependence of the robustness and quality of the generated multiphoton state on the externally applied magnetic field ($B$), integration time ($T_\mathrm{M}$), radiative time ($\tau_\mathrm{r}$) and the ratio between the electron and HH $g$-factors ($g_\mathrm{e}\over g_\mathrm{h}$). The model agrees very well with our experimental observations.  
%********************HERE*********************

\subsection{Characterization of the GHZ state}\label{sub1sec2}

%Figure4
\begin{figure*}
	\begin{centering}
		\includegraphics[width=1\textwidth]{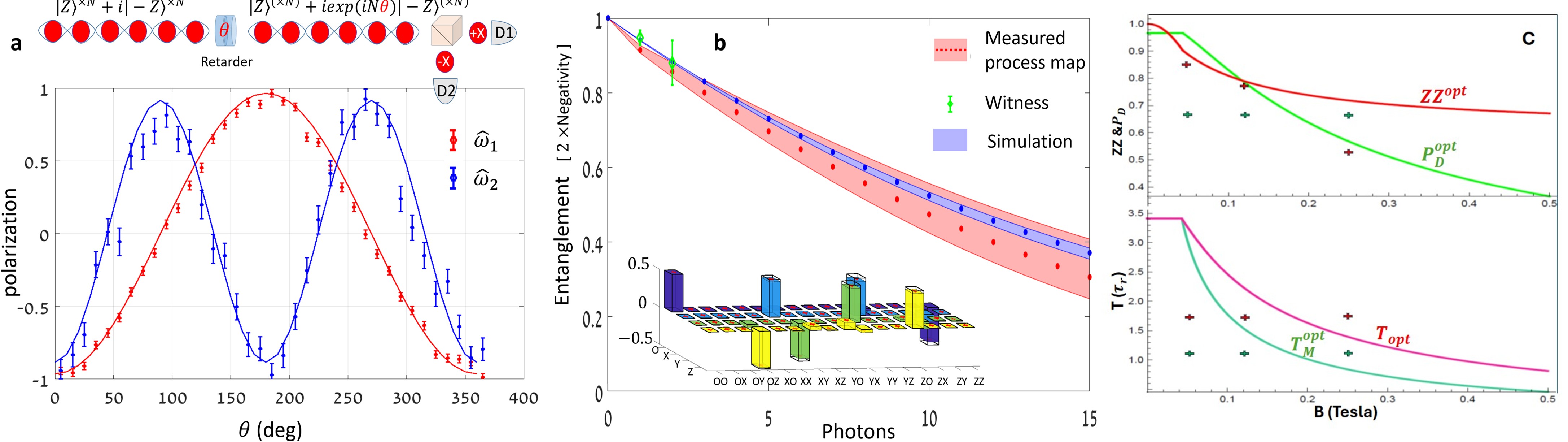}
		\par\end{centering}
	\caption{
		\label{fig: Measurement}
		{\bf Characterization of the multiphoton GHZ state.}
		({\bf a}) Upper panel: Schematic description of the experiment. Lower panel: Rectilinear degree of polarization ($\omega_n$-witnesses) of the multiphoton state as a function of the retarder's phase. 3-photon (red) and 4-photon (blue) correlation measurements of the witnesses $\hat\omega_1$ and $\hat\omega_2$, respectively.
		Points with error bars represent the measurements, and color-matched solid lines are best fits to
		$\hat\omega_n(\varphi)=-V_{\hat\omega_n}\cos(n\varphi)$, 
		where $n=1,2$ and $V_{\hat\omega_n}$ 
		are the fitted corresponding visibilities.
		({\bf b}) The localizable entanglement (LE) in the multiphoton GHZ state as a function of the number of photons in the state.
		The inset presents a full tomographic measurement of the periodic process map. The fidelity between the measured (colored bars) and ideal (empty bars) processes is $F=0.93$.
		The pink points and shaded area represent the LE deduced from the measured process map and its experimental uncertainty. 
		The blue points and shaded area represent the LE based on the model calculations \cite{Cogan2023} and the uncertainties in the QD measured parameters. 
		The green data points with error bars represent the estimations based on the visibility of the measured witnesses $\hat\omega_1$ and 
		%DAN PLEASE ADD COLOR GREEN
		$\hat\omega_2$ (Fig.~\ref{fig: Measurement}a).
		({\bf c}) Lower panel: The calculated optimal acquisition window $T_\mathrm{M}$ which maximizes the product $ZZ\times P_\mathrm{D}$ (green solid line) and the resulting $T_\mathrm{optimal}$ which nulls the average acquired phase (red solid line), as functions of the externally applied magnetic field. Color-matched data points present the experimental values that we actually used.
		Upper panel: The resulting optimized $ZZ$-witness (red solid line) and $P_\mathrm{D}$ (green solid line) as functions of the field. Color-matched data points present the experimental values.
	}
\end{figure*}

A frequently used way to characterize a multiphoton GHZ state is to measure the added optical phase of the photons upon passing a medium with known optical retardation. 
In principle, if the medium adds an optical phase of $\theta$, the polarization state of a linearly polarized classical light source (or a source of uncorrelated single photons), will rotate by $\theta$, while the polarization of an $N$-photon GHZ state will rotate by $N\theta$ \cite{Giovannetti2004,Peniakov2020}. 
Since the overall photon detection efficiency of the experimental setup enabled detection of up to 4 sequential photon events in a reasonable rate, we directly measured GHZ states of up to  $N=2$, which required, as explained above, detecting 4 sequential photon events.

In Fig.~\ref{fig: Measurement}a, we present by red (blue) data points the measured normalized degree of rectilinear polarization of $N=1$ ($N=2$)-all-photonic-GHZ-state, as a function of the optical phase $\theta$ acquired by passing through a LCVR. The measurements were performed with GHZ state produced at $B=0.05$ Tesla and $T_\mathrm{M} \approx 1.1\tau_\mathrm{r}$.
The red data points marked as $\hat\omega_1$-witness are obtained from sequential detection of 3 photon events and the blue data points marked $\hat\omega_2$-witness are obtained from sequential detection of 4 photon events. 
The measured degree of polarization is normalized in considerations with the imperfections in the HH-spin state initialization and final spin-projection.
The color-matched solid lines in the figure describe best fit to $\hat\omega_N(\theta)=-V_{\hat\omega_N}\cos(N\theta)$, where $N=1,2$ and $V_{\hat\omega_N}$ are the fitted corresponding visibilities. 
For an ideal $N$-multiphoton GHZ state one expects that $V_{\hat\omega_N}=1$ \cite{boto2000quantum} for any $N$.
In reality, however, for a non-ideal case $V_{\hat\omega_N}$ is expected to decay with increasing $N$. 
The characteristic decay rate of the visibility with the increase in $N$, can therefore be a measure for the robustness of the entanglement in the multiphoton states \cite{Peniakov2020}.
The expected phase multiplication between the $N=1$ and $N=2$ cases is clearly observed in Fig.~\ref{fig: Measurement}a and the visibilities of the measured witnesses are $V_{\hat\omega_1}=0.95\pm0.02$ and $V_{\hat\omega_2}=0.88\pm0.06$. 
These visibilities are marked as green data points in Fig.~\ref{fig: Measurement}b.
The entanglement characteristic decay length obtained by the ratio between these visibilities is given by $\zeta_{\hat\omega}=13\pm3$.

In Fig.~\ref{fig: Measurement}b we present the measured process map of the E-CNOT gate which generates spin-photon entanglement. 
The process map was obtained by performing full tomographic measurements at the optimal excitation rate  $\Gamma_\mathrm{exc}^\mathrm{opt}=1.46$ GHz, $B=0.05$~Tesla and $T_\mathrm{M}=1.1\tau_\mathrm{r}$.
The HH spin was initialized in six different basis states \cite{Cogan2020}. 
We applied the E-CNOT gate to each initialization and measured the resulting HH-photon density matrix by projections onto 24 polarization combinations (6 photon states multiplied by 4 spin states).
The physical process map that best fits these over-determined experimental measurements was then found using an edge-sensitive gradient-descent algorithm \cite{Cogan2022a}. 
The colored bars in Fig.~\ref{fig: Measurement}b describe the magnitude of the measured process map entries and the empty bars describe for comparison the expected magnitudes of an ideal CNOT gate entries. The fidelity between the measured and ideal gate is $0.93\pm0.01$. 

The process map is sufficient to fully characterize the entanglement robustness of an $N$-photon GHZ states by taking advantage of the periodic nature of the multiphoton state production \cite{Schwartz2016}. A spin-$N$-photon GHZ state is given by $\Phi^{\otimes N}(\hat{\rho}_\mathrm{h})$ where $\Phi$ represents the measured process map tensor and $\rho_\mathrm{h}$ is the initial HH spin density matrix. The $N$-photon spin state can be numerically formed with confidence levels that are determined by the experimental uncertainties in the measured process map, and the initial spin state. 

The robustness of the entanglement in the multiphoton state can be characterized by the localizable entanglement (LE) \cite{Popp2005}, defined here as twice the negativity \cite{Peres1996} in the density matrix of the first qubit $i=1$ and the last qubit $i=N$, while the polarizations of all the other qubits ($1<i<N$) are projected on 
the $\ket X$ base and measured.
For an ideal process map, LE equals unity for any $N$. In reality, however, LE decays exponentially with increasing $N$. 
The characteristic decay length $\zeta_\mathrm{LE}$ is a measure for the robustness of the entanglement in the generated multiphoton state. 

In Fig.~\ref{fig: Measurement}b we present by pink points the calculated LE of the $N$-photon GHZ states numerically generated by applying the measured process map $N$ times. 
The pink shaded area in Fig.~\ref{fig: Measurement}b represents one standard deviation uncertainty in the calculated LEs due to the uncertainty in the measured process map. 
The entanglement characteristic length obtained from these estimations is $\zeta_\mathrm{LE}^\mathrm{pm}=13_{-1}^{+2}$.

Likewise, the blue points in the figure represent calculations using the evolution model described in Ref.~\cite{Cogan2022}.
The shaded blue area reflects the experimental uncertainties in the measured QD parameters \cite{Cogan2022,Cogan2023}. The LE characteristic decay length according to this model $\zeta_\mathrm{LE}^\mathrm{model}=15\pm0.5$ agrees quite well with $\zeta_{\hat\omega}$ and with %the decay length 
that obtained from the measured process map ($\zeta_\mathrm{LE}^\mathrm{pm}$). 

\subsection{Optimization of the GHZ state production}\label{sub2sec2}
We discuss here the specific choice of the temporal window width $T_\mathrm{M}$, which affects the probability to detect the emitted photons ($P_\mathrm{D}$, Eq.~\ref{PdTm} in Methods), the entanglement robustness and the $ZZ$-witness, (Eqs.~\ref{phase_4}, \ref{phase_6}, in Methods) and the optimal excitation rate ($\Gamma_\mathrm{exc}^\mathrm{opt}$, Eq.~\ref{phase_5}, in Methods).
Since at a given field $B$, $P_\mathrm{D}$ increases with $T_\mathrm{M}$ while $ZZ$ decreases, we chose to define $T_\mathrm{M}^\mathrm{opt}$ which maximizes the product $P_\mathrm{D} \times ZZ$, (assuming $A_\mathrm{dec}=1$, since by Eq.  \ref{phase_6}, in Methods, for $T_\mathrm{M} > \tau_\mathrm{r}$ $A_\mathrm{dec}$ only very weakly depends on $T_\mathrm{M}$).

In the lower panel of Fig.~\ref{fig: Measurement}c we display the calculated $T_\mathrm{M}^\mathrm{opt}$ and the resulting excitation period $T_\mathrm{optimal}$ which nulls the average phase as functions of the externally applied field using green and red solid lines, respectively. Color-matched data points present the experimental values that we actually used.
Similarly, solid green (red) line in the upper panel of Fig.~\ref{fig: Measurement}c presents the resulted optimized value of $P_\mathrm{D}$ ($ZZ$) and 
the color-matched data points present the experimental values. 

\section{Discussion}\label{sec4}

Since $\Delta\varphi$ is proportional to the magnetic field (Eq.~\ref{phase_4}), at zero external magnetic field $\Delta\varphi$ vanishes, regardless of the size of $T_\mathrm{M}$. 
Therefore at vanishing external fields $T_\mathrm{M}$ can be extended up to $T_\mathrm{optimal}$ yielding $P_\mathrm{D}^\mathrm{max}$, without degrading the $ZZ$-witness. 
Therefore as clearly demonstrated in Fig.~\ref{fig: Measurement}c, the photonic GHZ generating device that we demonstrate here, performs best at vanishing or nearly vanishing external magnetic fields, 

In this demonstration we had to use minimal fields for initial spin preparation and final spin readout, both of which rely on timed spin precession.
The magnitude of this minimal field depends mainly on the hole spin dephasing time, which reflects the influence of the internal nuclear magnetic field \cite{Cogan2018} on the hole spin. 
Based on Fig.~\ref{fig: Measurement}c, the minimal field used here should still provide an excellent device for generating a robust multiphoton GHZ state, as indeed we demonstrated here 
with record entanglement length of 13 photons. 

The entanglement length is currently limited mainly by the decoherence of the electronic spin during the radiative lifetime, expressed as $A_{dec}$  in Eq. \ref{phase_6} in Methods. 
The main contribution to the fast electronic spin decoherence is its interaction with the fluctuations in the internal magnetic fields due to the QD nuclear spins.  
Using the approximation in Eq. \ref{phase_6}  and $\tau_{dec}\approx 1$ ns one gets $A_{dec} \approx 0.87$ and 
$\zeta_\mathrm{LE} \approx 13$ in agreement with our demonstration. 
Assuming a feasible 5-fold Purcell enhanced radiative rate \cite{Liu2018}, our approximation yields  $\Gamma_{exc}^\mathrm{opt}\approx 4$~GHz, $P_\mathrm{D} \approx 0.97$, and $ZZ\approx0.994$. The expected $\zeta_\mathrm{LE}$ at vanishing external magnetic field in this case is longer than 100 photons.

In summary, we demonstrate deterministic generation of a large-scale GHZ state of indistinguishable photons. 
For generating the state, we use repeated excitation of a semiconductor QD confined heavy-hole resulting in a string of single photons which are polarization entangled to the QD's spin. 
By adjusting the excitation rate and utilizing the opposite $g$-factors of the hole and the electron in its excited trion state, we demonstrate a natural spin echo which effectively increases the entanglement length of the GHZ state.  
We show that currently the robustness of the entanglement in the multiphoton state is about 13 photons, limited by the decoherence of the electron in the excited trion. 
We foresee further feasible improvement of the device by incorporating it within a wavelength-scale photonic microcavity, which will substantially reduce the radiative lifetime, while increasing the photon extraction efficiency. 
This in turn will dramatically increase also the generation rate, decrease the sensitivity to the internal magnetic fields and substantially increase the entanglement length.
This novel manner for producing large-scale GHZ states of indistinguishable photons in record high generation rate and entanglement robustness forms a perfect resource for constructing high-dimensional multiphoton states through probabilistic fusion \cite{paesani2023fusion,Rudolph2017,azuma2015all}.

\section{Methods}\label{sec3}

The time-dependent phase acquired by the trion (HH) can be written as $\varphi_\mathrm{t(h)}(t) =\omega_\mathrm{t(h)} t$. The angular frequency of the precessing trion (HH) spin around the magnetic field $B$ is given by $\hbar\omega_\mathrm{t(h)}=g_\mathrm{e(h)} \mu_\mathrm{B} B$, where $\hbar$ is the reduced Planck constant, $\mu_\mathrm{B}$ is the Bohr magneton, and $g_\mathrm{e(h)}$ is the electron (HH) Land\'e $g$-factor tensor component in the direction of the magnetic field $B$.
We define a temporal window of width $T_\mathrm{M} \le T$ after the excitation, during which we detect the emitted photon (see Fig.\ref{fig: Method}). 
The probability to detect the photon depends on this width: 
\begin{equation} 
	%P_D(T_M)= \int_{0}^{T_{M}}\frac{e^{-t/\tau_r}}{\tau_r}dt = 1-e^{-T_M/\tau_r}
	P_\mathrm{D}(T_\mathrm{M})= \int_{0}^{T_\mathrm{M}}\frac{e^{ -\frac{t'}{\tau_\mathrm{r}}}}{\tau_\mathrm{r}}dt' = 1-e^{-\frac{T_\mathrm{M}}{\tau_\mathrm{r}}},
	\label{PdTm}
\end{equation}
where $P_\mathrm{D}$ approaches determinism for $T_\mathrm{M} \gg \tau_\mathrm{r}$.
%The normalized probability that the trion decays at time $ 0 < t < T_{M}$ during the time window is $P(t)=\frac{1}{\tau_r(1-e^{-T_{M}/\tau_r})}e^{-t/\tau_r}$.

The average total phase acquired between two sequentially detected photons is then given by:
\begin{eqnarray}\label{phase_2}
	\langle\varphi\rangle &=& \frac{1}{P_\mathrm{D}(T_\mathrm{M})}\int_{0}^{T_\mathrm{M}}\frac{e^{ -\frac{t'}{\tau_\mathrm{r}}}}{\tau_\mathrm{r}}\varphi_\mathrm{h}(T-t')dt'\nonumber\\
	&+& \frac{1}{P_\mathrm{D}(T_\mathrm{M})}\int_{T}^{T+T_\mathrm{M}}\frac{e^{ -\frac{(t'-T)}{\tau_\mathrm{r}}}}{\tau_\mathrm{r}}\varphi_\mathrm{t}(t'-T)dt',
\end{eqnarray}
%Where $P(t)=\frac{1}{\tau_r(1-e^{-T_{M}/\tau_r})}e^{-t/\tau_r}$ is the normalized probability that the trion decays at time $ 0 < t < T_{M}$.
% $ \tau_r$ is the trion's radiative time, $T=\Gamma_{exc}^{-1}\ge T_M$ is the time difference between two sequential laser pulses, $\Gamma_{exc}$ the excitation repetition rate, and $\varphi_{t(h)}(t) =\omega_{t(h)} \cdot t$ is the trion (hole) time dependent acquired phase. The integration variable $t'$ represents the trion recombination time measured from the excitation at $t'\approx0$ (we neglect here the laser pulse width and the fast optical-phonon-assisted relaxation of the trion to its ground level, both of order 10 ps []). The measurement time window $T_M$ is chosen to optimize the product of the photon efficiency with the polarization witness \ref{suROP2024}. The angular frequency of the precessing trion (hole) spin around the magnetic field B is given by $\hbar\omega_{t(h)}=g_{e(h)} \mu_B B $, where $\hbar$ is the reduced Planck constant, $\mu_B$ is the Bohr magneton, and $g_{e(h)}$ is the electron (hole) effective Lande g-factor tensor component in the direction of the magnetic field B. 
where the first (second) term describes the contribution of the hole (electron) precession after (before) the detection of the first (second) photon. 

Eq.~\ref{phase_2} can be analytically solved yielding:
\begin{equation}\label{phase_3}
	%\langle\varphi\rangle=\omega_h T+(\omega_t-\omega_h)\tau_r(1+\frac{T_M/\tau_r}{1-e^{T_M/\tau_r}})
	\langle\varphi\rangle=\omega_\mathrm{h} T+(\omega_\mathrm{t}-\omega_\mathrm{h})\tau_\mathrm{r}\left(1+\frac{T_\mathrm{M}/\tau_\mathrm{r}}{1-e^{T_\mathrm{M}/\tau_\mathrm{r}}}\right)
\end{equation}
and for its standard deviation:
\begin{equation}\label{phase_4}
	%\langle\Delta\varphi\rangle^2=(\omega_\mathrm{t}^2+\omega_\mathrm{h}^2)\tau_r^2(1-\frac{(T_M/2\tau_r)^2}{sinh(T_M/2\tau_r)^2})
	\langle\Delta\varphi\rangle^2=(\omega_\mathrm{t}^2+\omega_\mathrm{h}^2)\tau_\mathrm{r}^2\left(1-\frac{(T_\mathrm{M}/2\tau_\mathrm{r})^2}{\sinh(T_\mathrm{M}/2\tau_\mathrm{r})^2}\right).
\end{equation}
For the generation protocol to work properly, the acquired average phase should vanish (or be an integer multiple of $2\pi$ \cite{Peniakov2020}).
By substituting $\langle\varphi\rangle=0$ in Eq. \ref{phase_3} one obtains the optimal excitation period ($T_\mathrm{optimal}$):
\begin{equation}\label{phase_5}
	T_\mathrm{optimal}=\frac{1}{\Gamma_\mathrm{exc}^\mathrm{opt}}=\tau_\mathrm{r}(1-\frac{g_\mathrm{e}}{g_\mathrm{h}})(1+\frac{T_\mathrm{M}/{\tau_\mathrm{r}}}{1-e^{T_\mathrm{M}/{\tau_\mathrm{r}}}}).
\end{equation} 
%As can be seen from Eq. \ref{phase_5} 
Eq. \ref{phase_5} states that for a given $T_\mathrm{M}$, the optimal excitation period, which nulls $\langle\varphi\rangle$, does not depend on the magnitude of the magnetic field. For a given QD ($\tau_\mathrm{r}, g_\mathrm{e}/g_\mathrm{h}$) it depends only sub-linearly on the window's width $T_\mathrm{M}\le T_\mathrm{optimal}$. 

By substituting the longest possible detection window $T_\mathrm{M}= T_\mathrm{optimal}$ into Eq.~\ref{phase_5}, one obtains an implicit expression for the maximal excitation period which nulls the average phase:
\begin{equation}\label{phase_5a}
	\frac{g_\mathrm{e}}{g_\mathrm{h}}=\frac{(xe^x-e^x+1)}{(x-e^x+1)}; x(\frac{g_e}{g_\mathrm{h}})=\frac{{T_\mathrm{optimal}^\mathrm{max}}}{\tau_\mathrm{r}}.
\end{equation} 
Note that $x(\frac{g_\mathrm{e}}{g_\mathrm{h}})$ is a monotonic function of its argument, that for large enough $|{g_\mathrm{e}\over g_\mathrm{h}}|$, it is approximately given by $|{g_\mathrm{e}\over g_\mathrm{h}}|+1$. 
The maximal photon detection probability $P_\mathrm{D}^\mathrm{max}$ is given by $P_\mathrm{D}(T_\mathrm{optimal}^\mathrm{max})=1-e^{-x}$ (Eq. \ref{PdTm}). 

By using the measured ratio between the electron and hole $g$-factors 
$g_\mathrm{e}\over g_\mathrm{h}$$=$$-0.367(4)\over{0.128(2)}$$=-2.87(5)$ \cite{Cogan2022}, 
we get $x=3.42(7)$ and $P_\mathrm{D}^\mathrm{max}\approx0.97$. Using the measured trion radiative lifetime $\tau_\mathrm{r}=0.4$~ns \cite{Cogan2022}, we obtain 
$T^\mathrm{max}_\mathrm{optimal} =x\tau_\mathrm{r}= 1.37(3)$ ns ($\Gamma^\mathrm{min}_\mathrm{exc}= 0.73(2)$~GHz). 

\begin{figure}
	\begin{centering}
		\includegraphics[width=1\columnwidth]{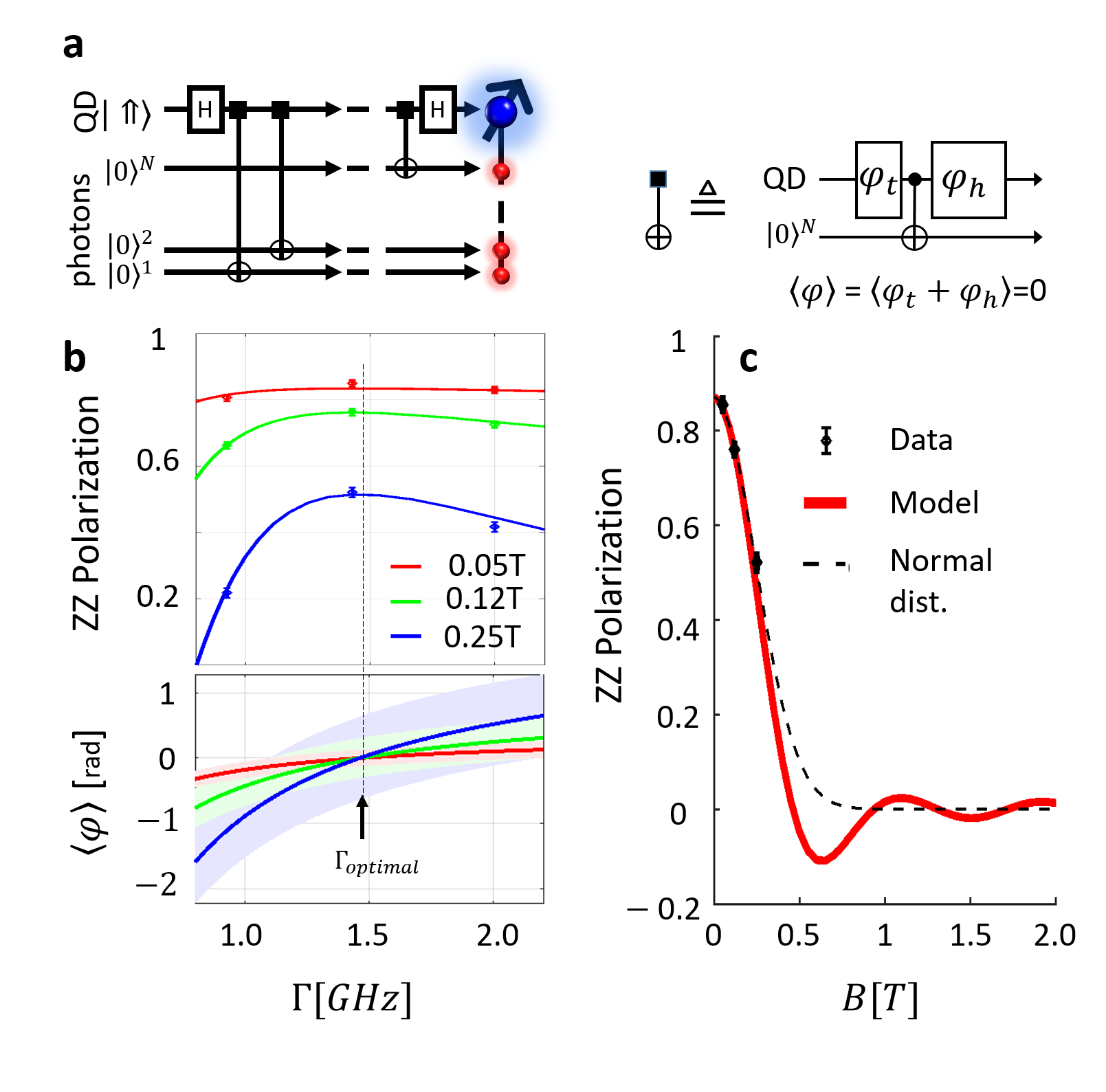}
		\par\end{centering}
	\caption{\label{fig: Method1} {\bf Optimization of the GHZ state generation.} 
		({\bf a}) A quantum circuit diagram describing the multiphoton GHZ state generation.
		({\bf b}) Lower panel: The calculated average accumulated phase $\langle \varphi \rangle$ during one cycle versus the excitation rate for various magnetic fields. The shaded areas represent one standard deviation $ \langle \Delta \varphi \rangle$. The black arrow indicates the field-independent optimal rate 
		at which $\langle\varphi \rangle=0$. %the two-qubit gate becomes an E-CNOT gate.
		Upper panel: The measured (data points) and calculated (color-matched solid lines) $ZZ$-witness 
		versus the excitation rate for various magnetic fields. 
		({\bf c}) The measured (black data points) and calculated (solid red line for the spin evolution model \cite{Cogan2022} and dashed black line assuming normally distributed $\varphi$) $ZZ$-witness versus the magnetic field magnitude. 
	}
\end{figure}

For this demonstration we chose $T_\mathrm{M}=0.43$~ns only. 
%by Eq.~\ref{phase_5} 
Our choice results in faster excitation rate $\Gamma_{exc}^\mathrm{opt}=1.46(2)$~GHz (Eq.~\ref{phase_5}) and lower detection efficiency $P_\mathrm{D}=0.66$
(Eq. \ref{PdTm}).		
In the lower panel of Fig.~\ref{fig: Method1}b, we plot using Eq.~\ref{phase_3} and Eq.~\ref{phase_4} the calculated $\langle\varphi\rangle$ and $\langle\Delta\varphi\rangle$ as functions of the excitation repetition rate for various externally applied in-plane magnetic fields. 
The optimal excitation rate which maintains $\langle\varphi\rangle=0$ 
 is marked by the vertical black arrow. 

%Figure 3

One of the indicators for the quality of the GHZ state is provided by the $ZZ$-witness defined as $\langle P_{Z}^{n}P_{Z}^{n+1}\rangle$ over every two sequential photons in the GHZ state, 
as defined in Eq. \ref{phase_NN+1}.  
The measured $ZZ$-witness is thus given by: 
\begin{equation}\label{phase_6}
	ZZ^{}=\langle A_\mathrm{dec} \cos (\varphi)\rangle \approx e^{-(\frac{\tau_\mathrm{r}}{\tau_\mathrm{dec}})^2} e^{-\frac{\langle\Delta\varphi\rangle^2}{2}}
\end{equation} 
where for the approximation in Eq. \ref{phase_6} we assumed for simplicity that 
i) $e^{-({\frac{\tau_\mathrm{r}}{\tau_\mathrm{dec}}})^2}$ ($\tau_\mathrm{dec}\approx 1$~ns) is a lower limit for $A_\mathrm{dec}$ resulting mainly from the electron spin interaction with the nuclear spins during the radiative cascade \cite{Cogan2018,Cogan2022, Merkulov2002, Bechtold2015} and ii) that $\varphi$ is normally distributed around zero.

In the upper panel of Fig.~\ref{fig: Method1}b we present the measured (data points) and calculated (color-matched solid lines, Eqs.~\ref{phase_4}, \ref{phase_6} and Ref.~\cite{Cogan2022}) $ZZ$-witness as a function of the repetition rate for the same magnetic field magnitudes as in the lower panel. 
Note that for all fields $ZZ$ is maximal at $\Gamma_\mathrm{opt}$.
The maximally obtained $ZZ$-witness values are $ ZZ_\mathrm{max}(B)=0.53\pm0.01$, $0.76\pm0.01$ and $0.85\pm0.01$, for $B=0.25, 0.12$ and $0.05$~T, respectively.
% The robustness of the entanglement increases with the decrease in the field, due to the linear dependence of the phase uncertainty (Eq. \ref{phase_4}). 
These values are presented in Fig.~\ref{fig: Method1}c as data points together with the model \cite{Cogan2022} calculations (solid red line) and the approximation from Eq.~\ref{phase_6} (dashed black line). 

At zero field, therefore, the device performs best. According to Eq. \ref{phase_6}, however, the extrapolated $ZZ_\mathrm{max}(B=0)=0.87(1)$ still deviates from unity mainly due to the electron spin dephasing induced by its interaction with the internal nuclear spins in the QD's vicinity. We note here, however, that according to this approximation, a feasible 5-fold Purcell induced reduction in the trion's radiative lifetime \cite{Liu2018} would lead to $ZZ_\mathrm{max}(B=0)=0.995$ (Eq.~\ref{phase_6}), very close to unity.

\section*{Acknowledgments} The support of the Israeli Science Foundation (ISF No. 1933/23 and ISF-NSFC No. 3094/23), the Shenzhen Municipal Funding Co-Construction Program (Project No. SZZX2301007), and the Guangdong Project (Grant No. 2025QN12X209) are gratefully acknowledged.

\bibliography{GHZcombinedZES}
\bibliographystyle{apsrev4-2}

\end{document}